# Exciton Dressing by Extreme Nonlinear Magnons in a Layered Semiconductor

**Authors:** Geoffrey M. Diederich[1§], Mai Nguyen[1], John Cenker[1], Jordan Fonseca[1], Sinabu Pumulo[1], Youn Jue Bae[2], Daniel G. Chica[3], Xavier Roy[3], Xiaoyang Zhu[3], Di Xiao[1,4*], Yafei Ren[5*], Xiaodong Xu[1,4*]

[1] Department of Physics, University of Washington, Seattle, WA 98195, USA

[2] Department of Chemistry, Cornell University, Ithaca, NY 14853, USA

[3] Department of Chemistry, Columbia University, New York, NY 10027, USA

[4] Department of Materials Science and Engineering, University of Washington, Seattle, WA 98195, USA

[5] Department of Physics, University of Delaware, Newark, DE 19716, USA

[§]Current affiliation: Department of Physics, University of Maryland Baltimore County, Baltimore, MD, 21250

*Correspondence to: Email: dixiao@uw.edu; yfren@udel.edu; xuxd@uw.edu

**Abstract**

**Collective excitations presenting nonlinear dynamics are fundamental phenomena with broad applications. A prime example is nonlinear optics, where diverse frequency mixing processes are central to communication, sensing, wavelength conversion, and attosecond physics. Leveraging recent progress in van der Waals magnetic semiconductors, we demonstrate nonlinear opto-magnonic coupling by presenting exciton states dressed by up to 20 harmonics of magnons, resulting from their nonlinearities, in the layered antiferromagnetic semiconductor CrSBr. We also create tunable optical side bands from sum- and difference-frequency generation between two optically bright magnon modes under symmetry breaking magnetic fields. Moreover, the observed difference-frequency generation mode can be continuously tuned into resonance with one of the fundamental magnons, resulting in parametric amplification of magnons. These findings realize the modulation of the optical frequency exciton with the extreme nonlinearity of magnons at microwave frequencies, which could find applications in magnonics and hybrid quantum systems, and provide new avenues for implementing opto-magnonic devices.**

**Main Text**

Magnonics is an emerging frontier that enables energy efficient information processing and transmission from spin waves[1]. It has gained considerable interest recently[2–5] due to the efficient coupling of magnons with disparate excitations, such as superconducting qubits, microwave photons, and phonons, which make magnons suitable as quantum transducers[6]. Recent advancements in 2D semiconducting magnets have provided a route to couple magnons to excitons, connecting the two quasiparticles separated by five orders of magnitude in energy[7,8], while higher order nonlinear excitations of quasiparticles in condensed matter systems have provided a fascinating route to controlled manipulation of materials properties. For example, nonlinearities in phononic systems has been shown to provide access to lattice control under excitation with ultrafast laser pulses[9]. In magnonic systems, nonlinearities are crucial for realization of magnon logic[10]. Thus, nonlinear magnonics[4,11–16] aims to combine the information carrying capacity of magnons with interactions and controls from strong nonlinearities. An unprecedented opportunity arises from the exciton-magnon coupling seen in 2D magnets as they may allow the mutual sensing or control afforded by nonlinear magnonics across many orders of magnitude in energy.

In this work, we report the observation of a series of optical sidebands generated via exciton dressing by nonlinear magnon modes in CrSBr. These nonlinear processes include high harmonic generation (HHG), sum frequency generation (SFG), difference-frequency generation (DFG), and parametric amplification (PA). The magnon-dressed exciton states originate from the unique combination of properties of CrSBr, a direct gap 2D van der Waals semiconductor that presents layered A-type antiferromagnetism[17,18]. In CrSBr, intralayer and interlayer magnetic coupling are ferromagnetic (FM) and antiferromagnetic (AFM), respectively, with spins aligned along the crystal *b* axis at equilibrium. Figure 1a shows the crystal structure and spin orientations at two distinct phases of the optical magnon precession, illustrated as arrows and separated by a dashed line. The magnon wavelength spans many unit cells and neighboring spins have nearly identical orientations, so the vertical dashed line in Fig. 1a can be thought of as a translation to a distant point in the crystal.

A recent report has demonstrated how the localization of the exciton wavefunction, $\psi_{Ex}$ in Fig. 1a, in CrSBr depends on its magnetic configuration[19]. This dependence leads to an exciton resonance energy shift proportional to the relative orientation of the spins in adjacent layers (angle $\varphi$ in Fig. 1b), i.e. $\mathbf{S}_1 \cdot \mathbf{S}_2$, where $\mathbf{S}_i$ is the magnetization in an individual layer. The strong coupling between the exciton resonance and interlayer spin alignment provides direct optical access to the magnons, where ultrafast optical pulses can generate and detect the magnons through transient modulation of exciton properties in pump-probe experiments. These measurements have led to reports of coherent magnon transport[7,20], hybridization of distinct magnon modes[8], and magnon coupled exciton-polaritons[21]. The ability to optically populate the magnons at high densities with ultrafast pulses allows us to generalize the opto-magnonic coupling to the regime of extreme magnon nonlinearity with optical read-out.

To measure the nonlinear magnonic processes, we performed transient optical reflectivity measurements of the exciton near the optical gap of 1.4 eV on exfoliated thin bulk CrSBr. The experimental temperature of 35 K is far below the Néel temperature of ~132 K[22]. Additional experimental details can be found in the supplementary information. The cartoon in Fig. 1c represents the nonlinear dependence of the magnon frequency, $\omega_m$, on the initial displacement of the spins from equilibrium, $\varphi_0$, in the perturbative nonlinear regime. With small spin displacements (orange line cuts), the system is well described by a simple harmonic oscillator. At larger spin displacements, illustrated by the blue and green lines, higher order processes, such as the second and third order harmonics shown in their corresponding interaction diagrams, will appear. In CrSBr, these magnon harmonics will modulate the exciton states. Measurement of these states are expected to give rise to optical side bands centered at the exciton resonance (Fig. 1c, right panel).

Figure 1d shows a typical transient optical reflectivity trace. The magnetic field $\mu_o H \approx 0.45$ T is applied along the crystal $a$ axis, resulting in a canted AFM configuration. The data has had the exponentially decaying electronic signal removed to emphasize the oscillations from the magnon-exciton coupling (see raw data in Extended Data Fig. 1). It is important to note in these data that the observed reduction in the oscillatory signal over time is not due to dephasing of the coherent magnons. The signal decay is indicative of coherent propagation of the magnons away from the localized measurement spot, as has been recently shown in work that estimates a minimum dephasing time of 5.3 ns.[20] The decay seen in our traces is consistent with the roughly 1 mm/ns group velocities measured in that study for thick (~100 nm) CrSBr flakes. Figure 1e presents the magnon spectrum, obtained through fast Fourier transformation (FFT) of the transient signal (see Extended Data Fig. 2 for $c$ axis field dependence). In this spectrum, the electronic response of the system appears at 0 Hz and there is a bright fundamental optical magnon sideband spanning from $\omega_o \sim 5$ GHz to $\sim 30$ GHz, as in prior reports[8,23,24]. Our new observation is the emergence of a sideband at $2\omega_o$ with a related field dependence to that of the optical mode, signifying the second harmonic generation (SHG) of the fundamental optical magnon.

To further support the SHG assignment of the $2\omega_o$ mode in Fig. 1e, we perform pump-fluence dependent measurements. Fig. 1f shows the dependence of the FFT amplitude of the $\omega_o$ (left) and $2\omega_o$ (right) modes under increasing pump excitation fluence for a pump beam with a ~1 μm radius. Here, the applied magnetic field is ~0.35 T along the $a$-axis. The extracted amplitude of the $\omega_o$ mode displays a linear power dependence until the magnon signal begins to saturate at ~450 μJ/cm$^2$. On the other hand, the amplitude of the $2\omega_o$ mode presents a clear quadratic power dependence up to the same saturation fluence. Since the bosonic magnon system doesn't saturate due to Pauli blocking, the saturation is likely due to local heating from the laser excitation. In the data shown in the remainder of this report, we keep the pump excitation below 300 μJ/cm$^2$ where any unwanted pump fluence-dependent effects are small. The fluence dependences in Figs. 1e and 1f confirm the nature of the $2\omega_o$ mode as SHG, where strong interactions of pump-generated magnons push the system beyond linear spin wave theory.

The observed SHG is a manifestation of the nonlinearity of the spin-wave dynamics that can be described by the Landau-Lifshitz-Gilbert (LLG) equations, which are inherently nonlinear. These dynamics, and the associated quadratic power dependence of the SHG mode, are indicative of perturbative harmonic generation. Perturbative expansions of the nonlinear dynamics can produce a few harmonics with quickly decaying intensity. In contrast, it is well known that in optical HHG, higher order harmonics scale non-perturbatively, as they arise from entirely different mechanisms[25–27]. As we will show later in this report, there is also non-perturbative magnon harmonic generation at play that produces HHG. This magnon HHG does not follow the picture presented in Fig. 1c or the perturbative expansion of the LLG equations.

We next seek to control the magnon nonlinear effects by exploiting the anisotropy of the system. Recent reports[8,23] have demonstrated that an in-plane off-axis magnetic field can break $C_2$ rotational symmetry in CrSBr. This allows them to hybridize and lifts the degeneracy between the optical and acoustic magnons at their crossing point, resulting in two magnon modes that both couple to the exciton. Specifically, it has been shown that applying fields at an angle, $\theta_{ab}$, in the plane of the crystal results in a splitting that is linearly proportional to $\theta_{ab}$ for angles smaller than ~15°. Therefore, we expect that nonlinear frequency mixing between the two modes will produce new magnon side bands beyond SHG. To measure this frequency mixing, we perform the same transient optical reflectivity experiments as described above, but with $\mu_o\mathbf{H}$ applied in the $ab$ plane at an angle $\theta_{ab} = 2°$.

Figure 2a shows the magnon spectrum resulting from the FFT of the transient optical measurements. For comparison, Fig. 2b is a simulated spectrum based on the LLG equation, where $\omega_1$ and $\omega_2$ represent the two magnon modes originating from the hybridization of the acoustic and optical magnon modes that present in-phase and out-of-phase precessions of the two spin sublattices, respectively. The hybridization arises from the symmetry breaking induced by magnetic fields (see the Methods section for simulation details)[8] and their relative spectral weight depends on the contribution of the optical magnon to the hybridized mode. In both the measured and simulated spectra, in addition to the SHG of the fundamental modes, two features are visible that are not integer multiples of either $\omega_1$ or $\omega_2$ but are their linear combinations (see Figs. 2c and 2d for a zoomed-in view of the data highlighting these peaks). At $\omega_1 + \omega_2$, there is a mode with weak field dependence sitting directly in the center of the avoided crossing of the SHG modes which corresponds to the SFG of the magnon modes. Close inspection of Fig. 2d appears to show an unexpected evolution of the spectral weight distribution with increasing magnetic field. This is caused by an additional change of the exciton resonance energy as the field is increased, resulting in an optical signal from the magnon that appears to have a slightly distorted field dependence. At $\omega_1 - \omega_2$, we see a strongly field dependent feature whose field dependence switches sign at the crossing point of $\omega_1$ and $\omega_2$, $\mu_o\mathbf{H} \approx 0.5$ T, corresponding to the DFG of the fundamental modes. As we will show, this DFG mode is highly tunable and can induce parametric amplification processes.

We can tune the SFG and DFG through the control of $\theta_{ab}$, which tunes the splitting between the $\omega_1$ and $\omega_2$ modes[8,23]. Using the DFG mode as an example, Fig. 3a shows the magnon spectrum as $\theta_{ab}$ is varied at fixed $\mu_o\mathbf{H} = 0.5$ T (time domain data in Extended Data Fig. 3). The small splitting between $\omega_1$ and $\omega_2$ in the data seen at $\theta_{ab} = 0°$ is likely due to imperfect sample mounting leading to a small tilt with respect to the in-plane magnetic field. As the splitting of the fundamental magnon modes increases with $\theta_{ab}$,[8] a low frequency DFG mode appears with its frequency increasing linearly with $\theta_{ab}$. Here, the disappearance of the DFG signal at frequencies lower than ~2 GHz is a result of our time domain background subtraction. We find that, as $\theta_{ab}$ is further increased, the DFG mode can be tuned into resonance with the lower branch ($\omega_2$) of the hybridized modes at $\theta_{ab} » 26°$. Extended Data Fig. 4 shows simulation results that qualitatively agree with the experimental results and tuning of the DFG mode, as well as a demonstration of the DFG mode at both positive and negative field angles.

Under the resonant condition, we see a small amplitude increase in the $\omega_2$ mode. This amplitude increase is due to parametric amplification of the $\omega_2$ magnon mode, similar to that recently reported in synthetic AFM systems[28]. Figure 3b shows energy level diagrams of two scenarios under which parametric amplification of the $\omega_2$ mode would occur. In the left panel, an $\omega_1$ magnon and an $\omega_2$ magnon, with arbitrary frequencies such that $\omega_1 > \omega_2$, are present in the system. In this scenario, the $\omega_2$ magnon induces the downconversion of the $\omega_1$ magnon by resonantly stimulating the emission of an $\omega_2$ magnon and, to conserve energy, another magnon at $\omega_1 - \omega_2$. However, the amplification is solely dependent on the $\omega_1$ and $\omega_2$ amplitudes and is not strictly tunable in this regime[28]. However, under the special condition $\omega_1 - \omega_2 = \omega_2$, we can see that the DFG magnon is resonant with $\omega_2$. This case is shown in the right panel of Fig. 3b. In this scenario, the downconversion of the $\omega_1$ magnon releases two additional $\omega_2$ magnons which can then stimulate further downconversion processes.

Figure 3c presents a higher resolution measurement near the crossing of the DFG and $\omega_2$ modes, where the parametric amplification is evident. The extracted frequencies of the DFG (black dots) and $\omega_2$ (red dots) magnons are shown in Fig. 3d, where the DFG frequencies are calculated from the difference of the $\omega_1$ and $\omega_2$ peaks. Further, Fig. 3e shows the extracted amplitudes of the $\omega_2$ and DFG peaks. We see that there is a distinct amplitude increase, about ~1.5 times amplification, in the $\omega_2$ magnon when the two modes are in resonance compared to the off resonant case. We can also compare this increase to the DFG amplitude measured away from their overlap region (black dashed line in Fig. 3e, measured at $\theta_{ab}$ » 22° and multiplied by 30 for scaling purposes). The linear sum of the two modes off resonance is much smaller than the mode amplitude on resonance, further supporting the observation of parametric amplification. The tunability of the nonlinear response offers a fascinating platform for magnon frequency conversion, generation of entangled magnons, and functional magnonic devices based on nonlinear interactions.

In addition to the rich nonlinear physics presented so far, we have found that CrSBr is also a remarkable material for studying extreme magnonic nonlinearities, such as HHG. With a sufficient signal-to-noise ratio, additional harmonic orders past SHG can be seen. Figure 4a presents a field-dependent magnon spectrum that shows a series of side bands at integer multiples of $\omega_o$. This harmonic series generates optical sidebands ranging from ~5 GHz to over 600 GHz. Figure 4b shows a linecut from this spectrum at a field of $\mu_o\mathbf{H} \approx 0.3$ T, presented in units of the fundamental magnon frequency, $\omega_o$, with the FFT background removed. In these data, an astonishingly high harmonic series can be seen, with the inset of Fig. 4b presenting harmonics beyond 20th order. The slope of the field-dependent magnon frequency exhibits clear changes across the higher-harmonics. To understand this behavior, we consider a magnon, with the fundamental frequency $\omega(B_a)$, and its $n^{th}$ harmonic, with frequency $n\omega(B_a)$. The slope of the fundamental frequency at any field value will be $\frac{\partial \omega(B_a)}{\partial B_a}$ and the slope of the $n^{th}$ harmonic will be $n\frac{\partial \omega(B_a)}{\partial B_a}$, resulting in an increase in the field dependence as the order increases.

The HHG processes seen above are repeatable across samples (see Extended Data Fig. 5 for HHG data from an additional sample). Notably, the higher harmonics show no appreciable increase in linewidth. As mentioned previously, our experimental decay is caused primarily by the coherent propagation of the magnons[20]. Thus, the relatively constant linewidths here suggest that the group velocities of the harmonics are not significantly different than that of the fundamental magnons. We have also checked that the harmonic orders here are not artifacts of the FFT or data processing, evidenced by the additional peaks in the avoided crossings of the higher harmonic bands. These peaks, like the SFG peaks discussed in Fig. 2, are caused by higher order nonlinear mixing processes ($a * \omega_1 + b * \omega_2$, where $a + b$ is equal to the order of the harmonic band). They can be seen in Extended Data Fig. 5 as a blurring of the high harmonic sidebands around $\mu_o\mathbf{H} \approx 0.5$ T. These peaks, which are not harmonic replicas of the individual bands but mixing processes, require nonlinear interaction in the system and confirm that the observed HHG is a real effect.

Until very recently, only the first few magnon harmonic orders were measurable[11,13,14] using conventional techniques such as Brillouin light scattering and ferromagnetic resonance spectroscopies. Recent experimental realization of magnon HHG have employed Fabry-Perot cavities[29] to achieve up to 17 orders of harmonic generation. More recently, resonant THz experiments have seen harmonic magnon generation up to 6$^{th}$ order[30]. These results underline the impact of our results. To our knowledge, only one report has exceeded the order of harmonics in our data. That work employed electron spin resonance (ESR) of NV centers in nanodiamonds[31] to measure many harmonics (> 50). However, due to the detection scheme employed in this study, signals were read out as subharmonics of the static ESR frequency, limiting the frequencies attainable to those less than the ESR frequency (~3 GHz). Further, the harmonics generated in that work were multiples of the drive frequency indicating a continuous spin wave dispersion regime. In our experiments, we generate dozens of harmonics of a discrete magnon mode and our frequency window is only physically limited by the time duration of our laser pulses, which could extend well into the THz range. For example, a typical Ti:Sapphire laser output of 100 fs pulses can resolve magnons up to a few THz in our measurements. Additionally, we use a pulsed excitation in our experiments, rather than a periodic drive. The high sensitivity of the optical measurements, enabled by the exciton-magnon coupling, allows us to use weak excitation pulses in our experiments ($< 300 \frac{\mu J}{cm^2}$). We expect that the use of amplified laser systems or engineered magnon cavities will more effectively access the nonlinearity, which could push the HHG further into the THz regime. This could allow coherent control of the magnetization and coupling to the known $A_{1g}^1$ phonon mode at ~3.6 THz[32]. The exciton-coupled nature of the magnons here also lead to exciting possibilities in quantum transduction, as CrSBr and the series of nonlinear magnons now present opportunities to couple excitations ranging across tens to hundreds of GHz to photons can be processed and measured using standard optical instruments.

The observed harmonic series in Fig. 4 is the magnonic analogue of optical HHG processes seen in atomic gases[25] and solids[26] that are highly useful for frequency conversion, measurement of crystal symmetries[33], and probing topological order[34]. This HHG indicates a deviation from the perturbative expansion of the LLG equations described earlier into the non-perturbative nonlinear regime. In contrast to the perturbative nonlinearity where the amplitude of high harmonics decays exponentially as the order increases, our results in Fig. 4b show that the amplitude presents a plateau after a decay in the first few orders. This is definitive evidence for the non-perturbative nature of the HHG. This is similar to non-perturbative optical HHG, where a plateau in the amplitude also forms beyond the first few orders.[35] However, in contrast to optical HHG in atomic gases that show only odd harmonics, we see both even and odd harmonics and the magnetic field does not affect which harmonics are generated.

The observation of even-order harmonics underscores the unique nature of magnonic HHG in comparison optical HHG, where only odd-order HHG is seen. In optical HHG, the process originates from the dynamics of electrons moving in a nonlinear potential depending on the electron coordinate ***x***. Due to the inversion symmetry of the atoms, this nonlinearity must be an even order of ***x***, which inherently leads to the absence of even-order harmonics. In contrast, our phenomenological model describes magnonic HHG arising from the nonlinear dynamics of the quasiparticles characterized by a generalized coordinate ***Q*** in configuration space. This ***Q*** can be even under the symmetry operations of solid media, allowing for both even and odd order nonlinearities in the potential energy function, which enables the generation of odd and even-order harmonics, respectively. The difference in amplitude between odd and even harmonics may reflect the relative strengths of the corresponding nonlinear terms for these quasiparticles. This nonlinear oscillator could be a higher energy magnon or phonon mode in the system. Extended Data Fig. 6 shows a simulated spectrum which qualitatively matches the experimental data (see the Methods section for the theoretical details). Note that in Fig. 4, there appears to be an enhancement of the HHG at low field. This apparent enhancement is due to field-induced excitonic shifts that modulate our experimental sensitivity and is not indicative of underlying physics.

Nonlinear magnonics holds the potential to become an important component in smaller, more energy efficient, and more robust[36] information systems than those based on electronic circuits[4]. We have shown that CrSBr is an excellent candidate system, as it displays a strongly nonlinear magnon response in the microwave range necessary for electronic measurement. The demonstrated unique optical access to the harmonic orders of the magnons facilitates magnon resonances over a much greater range and extends CrSBr potential as a quantum transducer[8,37]. In addition, the SFG and DFG magnons provide avenues for generating entangled magnons[38] and parametric amplification of weak magnon signals[28] with unprecedented tunability via the angle of an applied magnetic field. Finally, the observed modulated exciton states from the extreme nonlinear magnonic processes provides access to opto-magnonic physics and could open new possibilities for magnonic control of the excitonic response, or vice versa. The nonlinear response presented here, combined with the desirable energy range and optical coupling, further cements CrSBr as a promising candidate for implementing hybrid magnon information processing.

**Methods**
**Crystal growth:** The following reagents were used as received unless otherwise stated: chromium powder (99.94%, −200 mesh, Alfa Aesar), sulfur pieces (99.9995%, Alfa Aesar), bromine (99.99%, Aldrich), and chromium dichloride, (anhydrous, 99.9%, Strem Chemicals). To begin, CrBr3 was synthesized from (Cr: 1.78 g, 34.2 mmol and Br$_2$: 8.41 g, 52.6 mmol) with one end of the tube held at 1000°C and the other end at 50°C with a water bath. Of note, one end of the tube must be maintained below 120°C to prevent the tube from exploding from bromine overpressure. Chromium (0.174 g, 3.35 mmol), sulfur (0.196 g, 6.11 mmol), and CrBr$_3$ (0.803 g, 2.75 mmol) were loaded into a 12.7 mm O.D., 10.5 mm I.D. fused silica tube. The tube was evacuated to ~30 mTorr and flame sealed to a length of 20 cm. The tube was placed into a computer-controlled, two-zone, tube furnace. The source side was heated to 850°C in 24 h, allowed to soak for 24 h, heated to 950°C in 12 h, allowed to soak for 48 h, and then cooled to ambient temperature in 6 h. The sink side was heated to 950°C in 24 h, allowed to soak for 24 h, heated to 850°C in 12 h, allowed to soak for 48 h, and then cooled to ambient temperature in 6 h. The crystals were cleaned by soaking in a 1 mg mL−1 of CrCl$_2$ aqueous solution for 1 h at ambient temperature. After soaking, the solution was decanted and the crystals were thoroughly rinsed with DI water and acetone. Residual sulfur residue was removed by washing with warm toluene.

**Sample fabrication:** CrSBr crystals were exfoliated onto silicon wafers with a 90-nm silicon dioxide layer and thin (10s – 100s nm) flakes were identified by optical contrast. Though thin bulk flakes were used, the magnon has weak thickness dependence[7] and we saw no evidence of thickness dependent nonlinear effects in our measurements.

**Optical measurements:** Transient optical reflectivity measurements were performed by tuning the output of a titanium sapphire oscillator to 1.33 eV, overlapped with one of the CrSBr exciton states. The output was frequency doubled and the second harmonic and fundamental were separated into pump and probe arms of the experiment by a dichroic mirror. The probe beam was sent to a retroreflector mounted on a motorized translation stage in order to produce the pump-probe delay. Each beam was sent through a waveplate and polarizer to simultaneously attenuate the beams and align their polarization to the crystal axes. The beams were recombined and sent through a 0.6 NA microscope objective onto the sample. The back-reflected beam was measured on a photodiode with a lock-in amplifier demodulating the signal at the frequency of a mechanical chopper placed in the pump arm of the experiment. To produce the time domain data, the delay stage was continuously swept at low speed while streaming data from the lock-in amplifier to the host computer at a high sampling rate (> 100 kHz), which produced time traces with 400 femtosecond resolution in the data presented here. Multiple traces (4 < N < 25) were recorded and averaged as well as binned in the time domain, depending on the desired signal-to-noise ratio and frequency range of the spectra. The error bars in Figure 3 are the standard deviation of the extracted values over all N measurements, while the center is the average value. The samples were kept at 35 K, far below the Neel temperature of ~132 K[22], in an optical cryostat with an integrated vector magnet capable of applying fields up to 1 T along any arbitrary direction on the unit sphere. In order to determine the orientation of our applied field with respect to the crystal axes, a separate measurement was taken with the field amplitude set to the center of the avoided crossing while we varied only the field angle. Zero angle was then set to be the point where the splitting, $\Delta$, went to zero. Field angles were then set by appropriately varying the two orthogonal in-plane magnetic fields in our vector magnet.

**Simulation of nonlinear magnon spectra:**
CrSBr is a layered material with ordered spin where the intralayer coupling is FM while the interlayer coupling is AFM. In each layer, there are two magnetic atoms in the unit cell that form a square lattice. The nearest-neighboring magnetic atoms in two layers have a relative in-plane shift. In principle, one needs to consider two magnetic atoms per layer per unit cell and the exchange coupling between second or third nearest-neighboring sites may play a role. However, for simplicity, we consider only one magnetic atom per layer per unit cell and consider only the nearest neighboring exchange interaction. The validity of this simplified model is justified by previous comparison with experimental results.[8]
The Hamiltonian reads

$$H = J_2 \mathbf{S}_1 \cdot \mathbf{S}_2 - g\mu_B \mathbf{H} \cdot \sum_{i=1,2} \mathbf{S}_i + \sum_{i=1,2} -K_b (\mathbf{S}_i \cdot \hat{b})^2 + K_c (\mathbf{S}_i \cdot \hat{c})^2,$$

where $\mathbf{S}_{1,2}$ denotes the spin in layers 1 and 2, with $J_2 > 0$ for AFM interlayer coupling. The easy and hard axes are the b and c axes, respectively, with $K_{b,c} > 0$. $\mathbf{H}$ is the external magnetic field with $g$ and $\mu_B$ denoting the g factor and the Bohr magneton.

Here, we solve the full LLG equation written as the following

$$\dot{\mathbf{S}}_1 = -\mathbf{S}_1 \times \mathbf{H}_1$$
$$\dot{\mathbf{S}}_2 = -\mathbf{S}_2 \times \mathbf{H}_2,$$

with effective magnetic fields

$$H_1 = J_2 S_2 - g\mu_B H - 2K_b S_1 \cdot \hat{b}\hat{b} + 2K_c S_1 \cdot \hat{c}\hat{c}$$
$$H_2 = J_2 S_1 - g\mu_B H - 2K_b S_2 \cdot \hat{b}\hat{b} + 2K_c S_2 \cdot \hat{c}\hat{c}.$$

In the presence of a magnetic field along the $a$ direction, the spins tilt along the field direction and the tilting angle can be determined by $\partial H/\partial\phi = 0$. Here, the static $H$ is

$$\begin{aligned} H(\phi) &= J_2 S^2 \cos(\pi - 2\phi) - 2g\mu_B SH_a\sin\phi - 2K_b S^2\cos^2(\phi) \\ &= -J_2 S^2 \cos(2\phi) - 2g\mu_B SH_a\sin\phi - 2K_b S^2\cos^2(\phi) \end{aligned}$$

and

$$\begin{aligned} \frac{dH}{d\phi} &= 2J_2 S^2\sin(2\phi) - 2g\mu_B SH_a\cos\phi + 4K_b S^2\cos(\phi)\sin(\phi) \\ &= 2(J_2 + K_b)S^2\sin(2\phi) - 2g\mu_B SH_a\cos\phi \end{aligned}.$$

Thus,

$$\sin\phi = \frac{g\mu_B H_a}{2(J_2 + K_b)S}.$$

When the magnetic field deviates away from the $a$-direction, the tilting angles for spin 1 and spin 2 are different. Nevertheless, one can still find the equilibrium spin configuration by minimizing the total energy with respect to the tilting angles.

After the arrival of the pump pulse, the equilibrium spin configuration changes due to the transient excited electronic states. We model this effect by introducing a sudden change in the spin configuration from its equilibrium orientation by a small amount. When the magnetic field is along the $a$-direction, the system respects two-fold rotational symmetry about the $a$-axis. This guarantees a symmetric configuration after the arrival of the pump pulse. Therefore, our introduced change is chosen to respect this symmetry. When the magnetic field deviates from the $a$-axis slightly, this symmetry still approximately holds. We then numerically solve the LLG equation to obtain $S_{1,2}(t)$ as a function of time, $t$. $S_1(t)\cdot S_2(t)$ can then be obtained, whose change is proportional to the shift in the exciton energy. We then Fourier transform $S_1(t)\cdot S_2(t)$ to obtain the full spectrum of the magnon modes.

We introduce a phenomenological model to explain the HHG observed in our data. This model consists of an anharmonic oscillator that is bilinearly coupled to the coherently excited magnon. The physical nature of the anharmonic oscillator will be discussed later. It is described by the following Hamiltonian:

$$H = \frac{P^2}{2} + \frac{1}{2}\omega^2 Q^2 + \frac{v_3}{3}Q^3 + \frac{v_4}{4}Q^4,$$

where $P$ and $Q$ are the generalized momentum and coordinate of the oscillator, $\omega$ is its intrinsic frequency of the oscillator, and $v_{3,4}$ represent the nonlinear coupling coefficients within the oscillator. This nonlinearity differs from that what is in optical HHG. Models for optical HHG study the electron dynamics in real space, where the generalized coordinate represents the real-space coordinate that is odd under inversion symmetry. The inversion symmetry of the atomic potential ensures that the cubic-order nonlinearity vanishes. However, in our model shown above, $Q$ corresponds to the generalized coordinate for the normal mode of an elementary excitation that can be even under symmetry operations like inversion or two-fold rotation about the $a$-axis. Consequently, the symmetry permits both cubic and quartic order nonlinearities. This crucial difference enables our observation of even-order harmonics, which are absent in optical HHG.

The presence of coherent optical magnons, which exhibit strong oscillation over the ~ 3 nanosecond measurement period, generate a periodic driving force, $F(t)$, on the anharmonic oscillator. Its equation of motion reads

$$\ddot{Q} = -\omega^2 Q - v_3 Q^2 - v_4 Q^3 - \frac{\dot{Q}}{\Gamma} - F(t)$$

where $\Gamma$ is the damping coefficient and $F(t)$ oscillates at the optical magnon frequency $\omega_0$. When $v_3 = 0$, this equation reduces to the well-known Duffing equation, a canonical model for studying nonlinear physics. Although the magnons exhibit damping—leading to an exponential decay of the driving force—experiments have shown the magnon lifetime exceeds the measurement period.[20] Therefore, we assume a periodic driving force. The results are qualitatively the same with weak decay.

By numerically solving the above equation, we find that the motion of the anharmonic oscillator exhibits high harmonics of the driving frequency $\omega_0$. As shown in Extended Data Figure 6, the oscillator's dynamics display frequencies that are integer multiples of the driving frequency, including both even and odd harmonics up to approximately the 22nd order. The amplitudes of the Fourier components initially decrease, then plateau, and subsequently decrease again, which qualitatively agrees with the measured data. The linewidths of the peaks show a weak dependence on the harmonic order, also in agreement with experimental observations. The model reliably generates HHG phenomena across a wide range of parameters, provided that nonlinearity and driving are present. We note that a nonzero $v_3$ is essential in generating even harmonics. By tuning the relative amplitude of $v_3$ and $v_4$, one can control the relative amplitudes of even and odd order harmonics, which is related to the even-odd oscillation of the amplitudes of the high harmonics in Fig. 4b.

This phenomenological explanation relies on the assumption that 1) the anharmonic oscillator is bilinearly coupled to the optical magnon, and 2) the anharmonic oscillator also couples to the exciton. Since the ultrafast pump pulse excites electronic excitations instantaneously, it is expected to suddenly modify the spin interactions and anisotropy, enabling a bilinear coupling between the optical magnon and other high frequency magnons. These magnon modes can also couple to the exciton as they alter the relative spin angles. Alternatively, the anharmonic oscillator could also represent phonons, such as interlayer shear phonons, which can modify spin-exchange pathways and thus couple to magnons. These phonons can exhibit nonlinearity and couple to excitons as well. It is possible that both processes contribute to the HHG.

While the phenomenological model produces results that qualitatively agree with our experiments, the microscopic mechanism remains elusive. Moreover, there are alternative mechanisms for generating HHG, like inhomogeneity in soft ferromagnetic thin films with dipole-dipole interaction under a periodic driving. To elucidate the microscopic origin of the HHG quantitatively, further experiments and theoretical calculations are required, which we leave for future work.

**Data Availability**

The datasets generated during and/or analyzed during this study are provided with this paper. All other data that support the plots within this paper and other findings of this study are available from the corresponding author upon reasonable request.

Figures

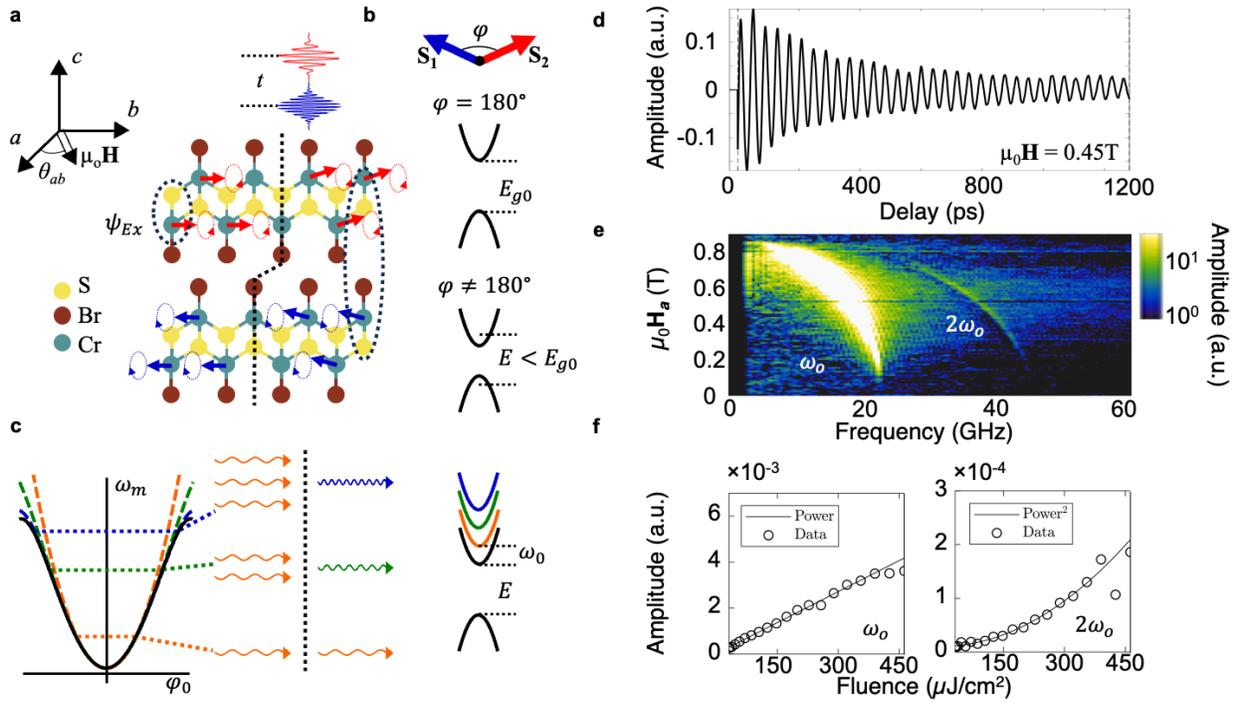

**Fig. 1 | Coupling of excitons and nonlinear magnons in CrSBr. a,** Crystal structure of CrSBr with two distinct spin orientations corresponding to distinct phases in the optical magnon precession. $\theta_{ab}$ denotes the angle between the in-plane magnetic field and the intermediate magnetic axis $a$. Black dashed ovals illustrate the delocalization of the exciton wavefunction, $\psi_{Ex}$, at two different points in the spin wave where pump and probe pulses excite and measure the system at some delay, $t$. **b,** Schematic description of the exciton dependence on interlayer spin alignment in CrSBr, where j represents the angle between the magnetization orientation of adjacent layers and $E_{g0}$ is the exciton energy in the layered AFM state. **c,** Cartoon showing the nonlinear dependence of the magnon frequency, $\omega_m$, on the initial angle between spins in adjacent layers, $j_o$, and its effect on the exciton in the perturbative regime. Dashed curves show approximations of varying orders. Horizontal dotted lines show regions where the behavior of the system is described by the various approximations and are connected to interaction diagrams for harmonic generation processes. The right cartoon depicts the resulting magnon-dressed exciton states. **d,** Transient optical reflectivity trace from thin CrSBr at $\mu_oH \approx 0.45$ T. **e,** Magnon spectrum obtained from the fast Fourier transform of the transient data. **f,** Power dependence of the FFT amplitudes of the fundamental ($\omega_o$) and second harmonic ($2\omega_o$) magnon modes measured at $\mu_oH \approx 0.35$ T. The magnetic field is applied along the $a$ crystal axis for the data in **(d-f)**.

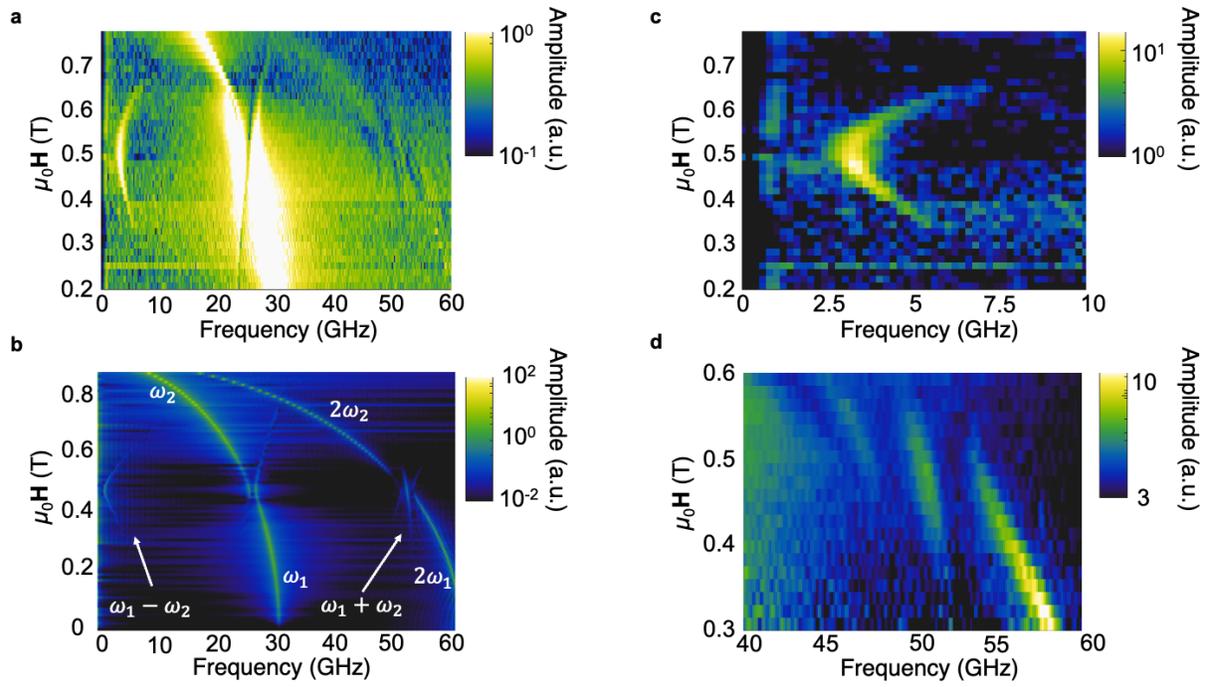

**Fig. 2 | Nonlinear coupling of coherently hybridized acoustic and optical magnon modes. a,** Measured magnon spectrum with $\theta_{ab} = 2°$. **b,** Simulated magnon spectrum with $\theta_{ab} = 2°$, containing only nonlinear terms. **c – d,** Measured magnon spectra with axes scaled to highlight the optical side band resulting from the exciton coupling to the difference-frequency (**c**) and sum-frequency (**d**) generation between the hybridized magnon modes.

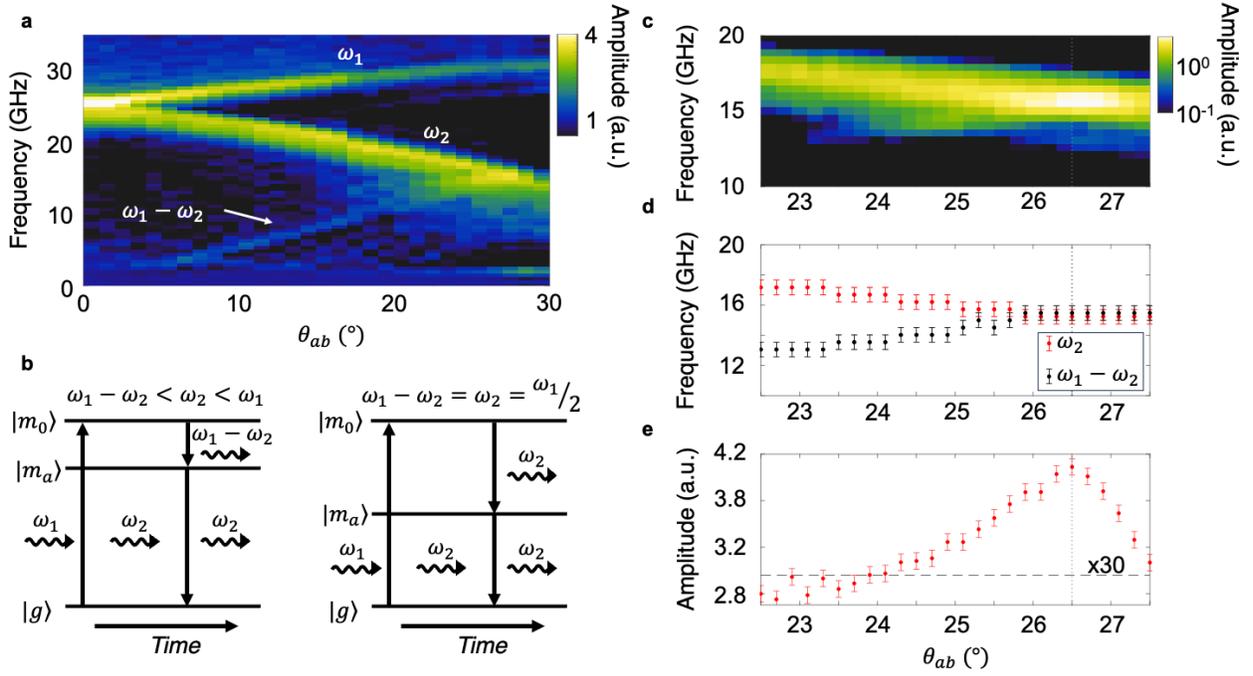

**Fig. 3 | Tunable DFG and parametric magnon amplification in CrSBr. a,** Measured magnon spectrum with respect to magnetic field angle $\theta_{ab}$. **b,** Schematic representation of the PA process. See text for details. **c,** Measured magnon spectrum focused on the region of overlap between the DFG and $\omega_2$ magnon modes. **d,** Frequencies of the $\omega_2$ (red) and DFG (black) modes extracted from (**c**). The standard deviation of the peak locations was smaller than our pixel size, which is set as an upper bound on the fitting error. **e,** Amplitude of the $\omega_2$ peak extracted from (**c**), where error bars signify the standard deviation between 25 measurements. The black dashed line indicates the amplitude of the DFG mode (multiplied by 30) when the two peaks are not overlapped.

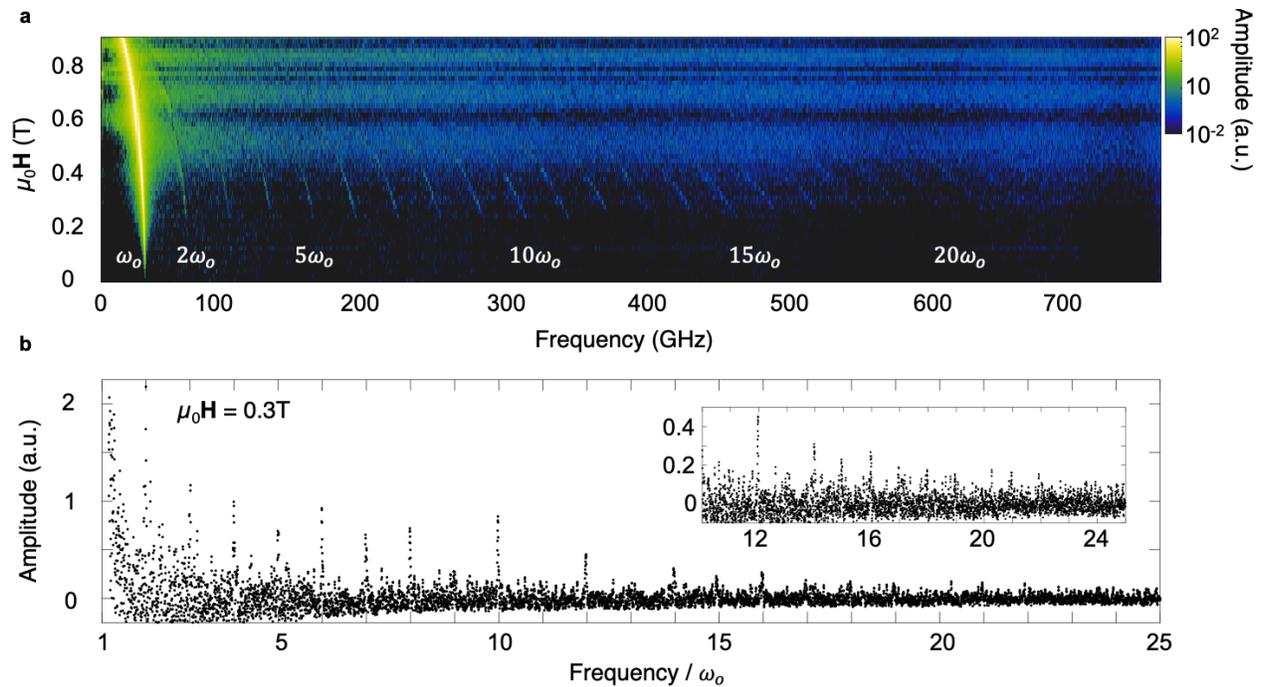

**Fig. 4 | Magnon high harmonic generation. a,** Field-dependent magnon spectrum presenting high harmonic magnon modes. **b,** Linecut taken from (**a**) at $\mu_o H \approx 0.3$ T, presenting harmonics beyond 20$^{th}$ order. Inset shows harmonics of 11$^{th}$ order and higher with the axes rescaled. In this figure, the fundamental magnon extends past the vertical scale to an amplitude ~100.

**Acknowledgments**

The authors thank Liang Fu, Mark Rudner, and Gil Refael for helpful discussions. This work was mainly supported by the Department of Energy, Basic Energy Sciences, Materials Sciences and Engineering Division (DE-SC0012509). Sample fabrication and optical measurements are partially supported by AFOSR FA9550-19-1-0390 and FA9550-21-1-0460. Synthesis of the CrSBr crystals is supported by the NSF MRSEC on Precision-Assembled Quantum Materials (DMR-2011738) and Programmable Quantum Materials, an Energy Frontier Research Center funded by the U.S. Department of Energy (DOE), Office of Science, Basic Energy Sciences (BES), under award DE-SC0019443. XX acknowledges support from the State of Washington funded Clean Energy Institute and from the Boeing Distinguished Professorship in Physics.## Author Contributions

XX, DX, YR, and GD conceived the project. GD performed the measurements with help from MN, SP, JC, and JF. MN, SP, and JC fabricated the samples. GD, MN, JC, YB, XYZ, YR, DX, and XX analyzed the data and interpreted the results. YR and DX built the model and performed the simulations. DGC and XR grew the CrSBr crystals. GD, MN, XYZ, YR, DX, and XX wrote the manuscript with input from all authors. All authors discussed the results.

## Competing Interests

The authors declare no competing financial interests.

# Extended Data for: Exciton Dressing by Extreme Nonlinear Magnons in a Layered Semiconductor

Geoffrey M. Diederich, Mai Nguyen, John Cenker, Jordan Fonseca, Sinabu Pumulo, Youn Jue Bae, Daniel G. Chica, Xavier Roy, Xiaoyang Zhu, Di Xiao, Yafei Ren, Xiaodong Xu

Corresponding authors: dixiao@uw.edu; yfren@udel.edu; xuxd@uw.edu

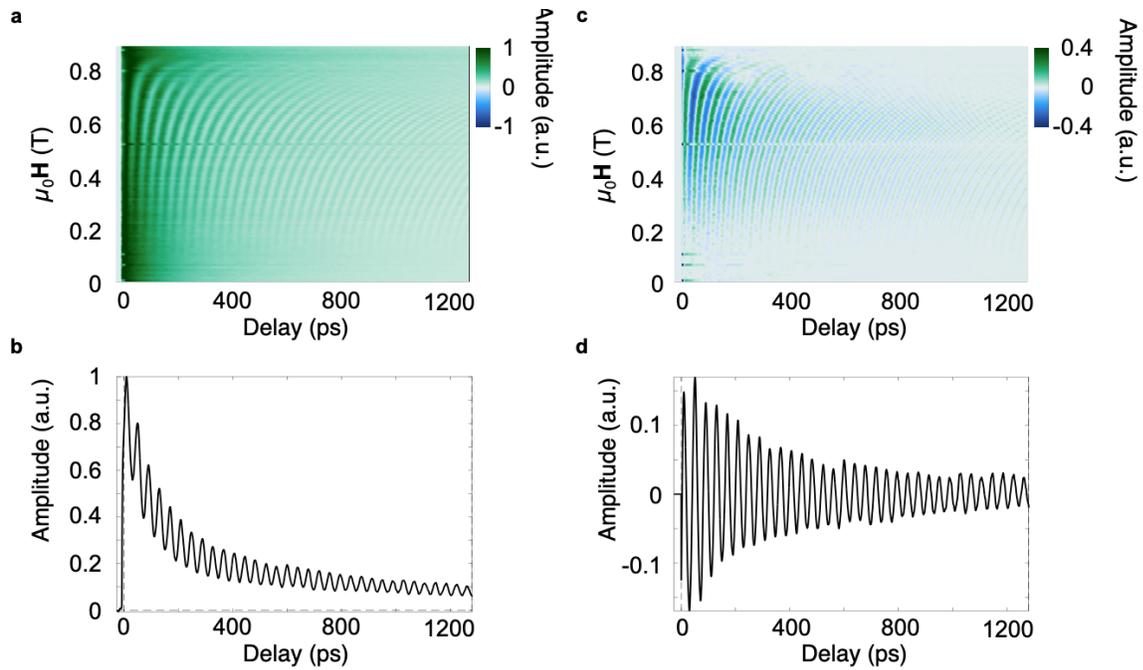

**Extended Data Fig. 1. Field dependent transient reflectivity data. a,** Raw transient reflectivity data corresponding to the spectrum presented in Fig. 1d. **b,** Linecut taken from (a) at $\mu_oH \approx 0.45$ T. **c,** Same data as in (a) with the exponential decay removed. **d,** Linecut taken from (c) at $\mu_oH \approx 0.45$ T.

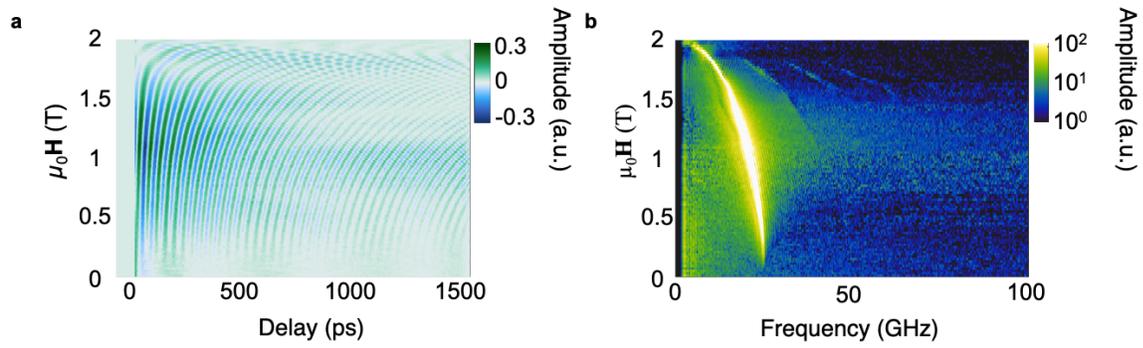

**Extended Data Fig. 2. Magnon HHG under *c* axis field. a,** Transient reflectivity data from CrSBr under a field applied along the *c* (hard) crystal (magnetic) axis. **b,** Corresponding magnon spectrum presenting the first few harmonic orders.

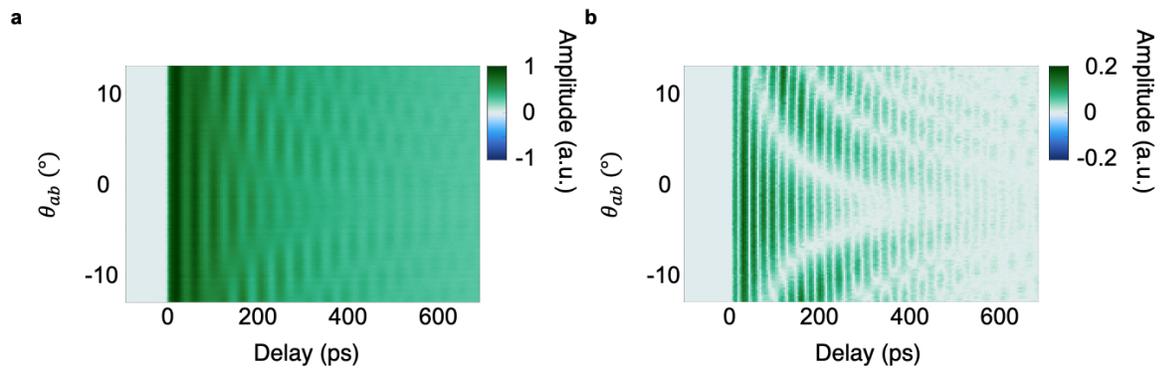

**Extended Data Fig. 3. Field angle dependent transient reflectivity data. a,** Raw transient optical reflectivity data corresponding to the spectrum presented in Fig. 3a. **b,** Same data as in (a) with the exponential decay removed.

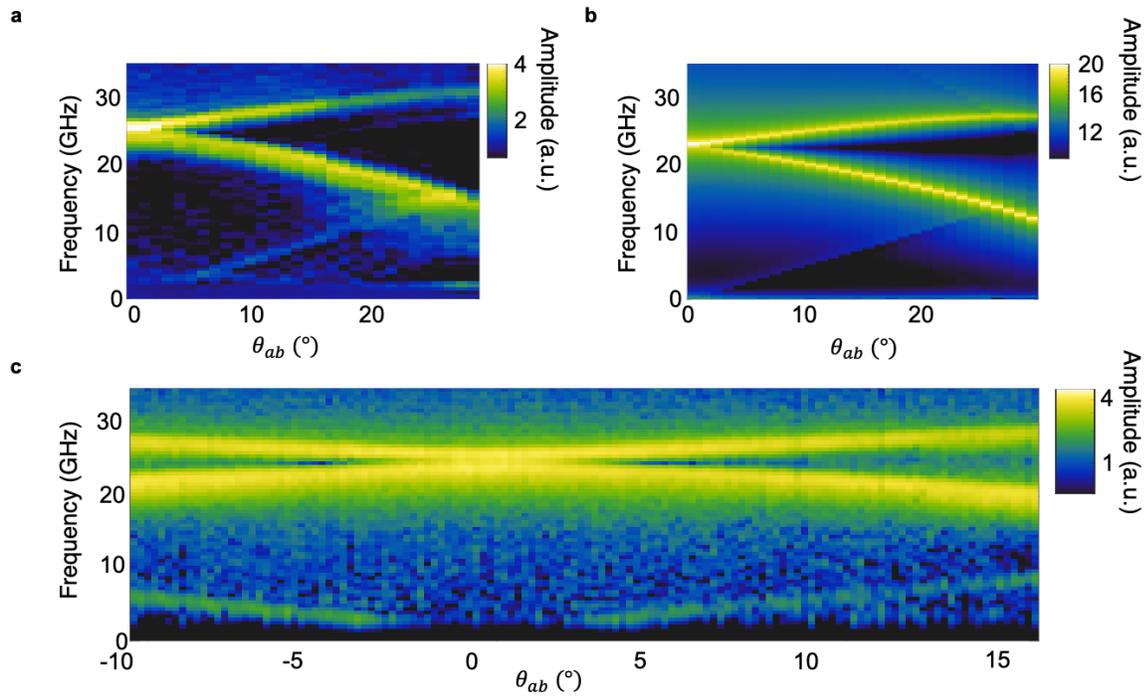

**Extended Data Fig. 4. Comparison of field angle dependent data with simulations**. **a,** Same measured magnon spectrum presented in Fig. 3b. **b,** Simulated magnon spectrum. **c,** Additional dataset showing the DFG mode at positive and negative field angles.

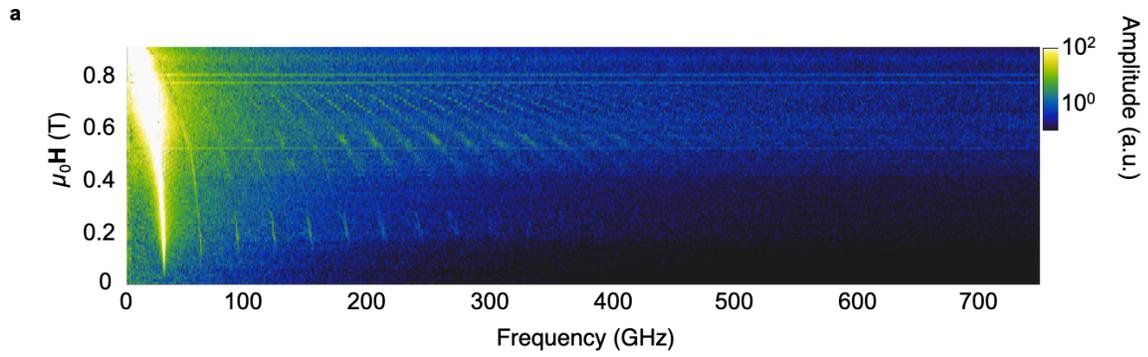

**Extended Data Fig. 5. HHG spectrum from an additional CrSBr sample**. **a,** Magnon spectrum presenting HHG from a ~90 nm thick flake, roughly half the thickness of the sample shown in the main text.

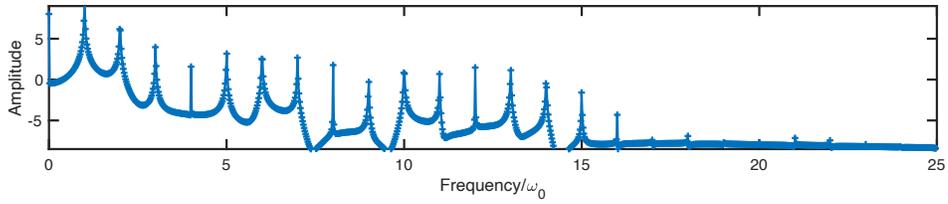

**Extended Data Fig. 6. Model calculation of HHG.** The amplitude of the Fourier components as a function of frequency is shown. The amplitude first decays, then reaches a plateau, and then decays again. Both even and odd harmonics exist. The model parameters used to produce this spectrum are $\omega = 1 = 10\omega_0$, $v_3 = 1$, $v_4 = \frac{1}{2}$, $\Gamma = 10$, and $F(t) = 0.5\sin(\omega_0 t)$.